# Comparison between two methods of post-Newtonian expansion for the motion in a weak Schwarzschild field

**M. Arminjon**

Laboratoire "Sols, Solides, Structures"
[CNRS / Université Joseph Fourier / Institut National Polytechnique de Grenoble]
B.P. 53, 38041 Grenoble cedex 9, France. E-mail: arminjon@hmg.inpg.fr

**Abstract.** The asymptotic method of post-Newtonian (PN) expansion for weak gravitational fields, recently developed, is compared with the standard method of PN expansion, in the particular case of a massive test particle moving along a geodesic line of a weak Schwarzschild field. First, the expression of the active mass in Schwarzschild's solution is given for a barotropic perfect fluid, both for general relativity (GR) and for an alternative, scalar theory. The principle of the asymptotic method is then recalled and the PN expansion of the active mass is derived. The PN correction to the active mass is made of the Newtonian elastic energy, augmented, for the scalar theory, by a term due to the self-reinforcement of the gravitational field. Third, two equations, both correct to first order, are derived for the geodesic motion of a mass particle: a « standard » one and an « asymptotic » one. Finally, the difference between the solutions of these two equations is numerically investigated in the case of Mercury. The asymptotic solution deviates from the standard one like the square of the time elapsed since the initial time. This is due to a practical shortcoming of the asymptotic method, which is shown to disappear if one reinitializes the asymptotic problem often enough. Thus, both methods are equivalent in the case investigated. In a general case, the asymptotic method seems more natural.

**1. Introduction**

Schwarzschild's first exact solution to the Einstein field equations provides a description of the space-time metric outside a spherically symmetrical object. It is often used to model the motion of a planet around the Sun, as the geodesic motion of a mass point in a Schwarzschild field ("Schwarzschild's motion"). The relevant field is that produced by the Sun, when the latter is assumed spherically symmetric and isolated. In particular, for Mercury, Schwarzschild's motion produces an advance in perihelion of 43" per century, as compared with the motion predicted with only the "Newton-like" part of the field. Since great astronomers: Le Verrier, then Newcomb, and still later Clemence, found that Newton's theory leaves an unexplained advance in Mercury's perihelion, which is precisely 43" per century according to Clemence, this is a great success of Einstein's general relativity (GR) – "by far the most important experimental verification of general relativity", according to Weinberg [1, p. 198]. The construction of modern ephemerides for the solar system does include corrections of GR, either directly in the form of standard post-Newtonian (PN) equations of motion for a system of mass points [2], or as analytical complements to the Newtonian motions in the solar system, based on the standard PN approximation of Schwarzschild's solution for the Sun [3]. Both methods give nearly identical results [3], hence it is the correction due to Schwarzschild's motion in the field of the Sun that produces the standard effects of GR on the main bodies of the solar system. In other words, the PN corrections due to the planets, as expressed in the standard (Einstein-Infeld-Hoffmann) PN equations of motion, are negligible, or nearly so, at the present observational accuracy.

However, besides the *standard* method of PN expansion [1, 4-5], whose resulting equations were used in the works quoted above, there is now a different method of expansion for weak gravitational fields. The new method consists in applying the usual method of asymptotic expansion for a system of partial differential equations (see *e.g.* Kevorkian & Cole [6]), therefore we shall call it the *asymptotic* method of PN expansion. This method leads, in particular, to expanding all fields with respect to the small parameter, whereas only the gravitational field is expanded in the standard method. The asymptotic method has been initiated by Futamase & Schutz [7], with new mathematical developments given by Rendall [8]. To the author's knowledge, the results obtained so far with this method for extended bodies in general relativity (GR) are limited to the local equations of motion and do not include the derivation of global equations of motion for the mass centers. Such global equations have been obtained [9-10] within an alternative, scalar theory of gravitation, by an integration of the local equations which were previously derived [11] using the asymptotic method in this theory. This involved long calculations, and the equations obtained include a number of terms. It would be even more so in GR, a tensor theory. But there is one important case where the application of the asymptotic method is easy, namely the case of one spherical body with test particles orbiting around it. This case is important because, as recalled above, it gives the overwhelming part of the standard PN corrections in the solar system. It thus seems appealing to attempt a precise comparison between the standard and asymptotic methods in that case, and this is the aim of this paper.

*Section 2* contains the body of the paper, and is structured as follows. It is first noted that, for a perfect fluid, the active mass in Schwarzschild's solution contains both rest mass and internal energy. Then the principle of the asymptotic method of PN expansion is recalled and it is shown that the rest mass and internal energy come into play at different approximation levels. Since, in the static case with spherical symmetry, the alternative scalar theory also predicts Schwarzschild's exterior solution with geodesic motion, the PN expansion of the active mass is given both for GR and for the scalar theory. Afterwards two different equations of motion are derived: a "standard" one and an "asymptotic" one. Section 2 ends by the numerical illustration for Mercury, when it is modelled as a test particle in the Sun's gravitational field. *Section 3* presents our conclusions.

**2. Post-Newtonian approximations to the Schwarzschild motion for a mass point**
*2.1 The active mass in Schwarzschild's exterior solution*

The usual form of Schwarzschild's exterior solution is:

$$ds^2 = \left(1 - \frac{2GM}{c^2 r}\right) c^2 dt^2 - \frac{1}{1 - \frac{2GM}{c^2 r}} dr^2 - r^2 d\Omega^2, \quad d\Omega^2 \equiv d\theta^2 + \sin^2\theta \, d\phi^2, \qquad (1)$$

where $c$ is the velocity of light, $G$ is the constant of gravitation, and $M$ is the "active mass". In GR, there is no *a priori* geometry, and the coordinates $t$, $r$, $\theta$ and $\phi$ gain their meaning precisely through the expression (1) of the metric. In the alternative scalar theory, $t$ is the "absolute time", while $r$, $\theta$ and $\phi$ are spherical coordinates centered at the symmetry centre, related to Cartesian coordinates ($x^i$) for the flat "background space metric" **g**$^0$ by the usual transformation:

$$x^1 = r\sin\theta\cos\phi, \quad x^2 = r\sin\theta\sin\phi, \quad x^3 = r\cos\theta. \qquad (2)$$

In GR, one may also define a new spatial coordinate system ($x^i$) from the starting system ($r$, $\theta$, $\phi$) by Eq. (2), of course. (The time coordinate $t$ is left unchanged in this transformation.)

The expression of the active mass $M$ depends on both the theory and the constitutive law assumed for the material making the massive spherical body B. In GR, this expression is known for a perfect fluid [12-13], it is

$$M = 4\pi \int_0^R r^2 \tau(r)\,dr = \int \tau\,dV, \quad dV \equiv dx^1\,dx^2\,dx^3, \tag{3}$$

where $R$ is the radius of the body (such that $\tau(r) = 0$ for $r > R$), and where

$$\tau = \mu^* \quad \text{(GR)}, \tag{4}$$

with $\mu^*$ the proper mass-energy density, such that the energy-momentum tensor of the fluid writes

$$T^{\mu\nu} = (\mu^* + p/c^2)\,U^\mu U^\nu - (p/c^2)\,\gamma^{\mu\nu} \tag{5}$$

($p$ is the pressure, $\mathbf{U} = (U^\mu)$ is the 4-velocity, and $(\gamma^{\mu\nu})$ is the inverse matrix of the component matrix $(\gamma_{\mu\nu})$ of the space-time metric tensor $\boldsymbol{\gamma}$). Moreover, $\mu^*$ is the sum of the proper rest-mass density and the proper volume density of the (elastic) internal energy:

$$\mu^* \equiv \rho^*(1 + \Pi/c^2). \tag{6}$$

To be definite, we consider a barotropic fluid, such that the local value of the proper rest-mass density depends only on the local pressure: $\rho^* = F(p)$. Then the internal energy also depends on the pressure only [4]:

$$\Pi = P(p), \quad P(p) \equiv \int_0^p \frac{dq}{F(q)} - \frac{p}{F(p)}, \tag{7}$$

hence the field $p$ determines all matter fields (the velocity field is zero here). In the scalar theory, Eq. (3) holds true [14] but, instead of Eq. (4), the source of the gravitational field is $\tau = (T^{00})_E$ in the most general case,[1] hence for a perfect fluid, by Eq. (5):

$$\tau = \left(\mu^* + \frac{p}{c^2}\right)\frac{\gamma_v^{\,2}}{f} - \frac{p}{c^2 f} \tag{8}$$

[10], with $\gamma_v$ the Lorentz factor, which is 1 here (static case). Hence, in contrast with Eq. (4) valid for GR,

$$\tau = \mu^*/f \quad \text{(scalar theory)}; \tag{9}$$

and $f$ is the scalar gravitational field, *i.e.* the (0,0) component of the space-time metric in the preferred frame: $f = (\gamma_{00})_E$.[1] In the spherical static case, we have [14]:

$$f(r) = 1 - 2\,V(r)/c^2, \quad V(r) \equiv \int_r^\infty \frac{GM(u)}{u^2}\,du, \tag{10}$$

where

$$M(r) \equiv 4\pi\int_0^r u^2 \tau(u)\,du. \tag{11}$$

---

[1] Index E means that $T^{00}$ (and also $\gamma_{00}$) is taken in coordinates bound to the preferred frame E, and with $x^0 = ct$ as the time coordinate, where $t$ is the "absolute time" (the preferred time of the theory).

(Hence, Eq. (3) gives $M = M(R)$ only in an implicit way for the scalar theory. Moreover, in this theory, a static situation is obtained only if $\tau$ is time-independent *in the preferred frame.*)

*2.2 Asymptotic expansion of the active mass*

Let us look for an approximate expression of $M$ in the case that the gravitational field (which is produced by B) is weak. The latter situation may be grossly defined, in general, by the vague assumption that the space-time metric differs little from a Galilean metric. In the present case, the departure of the metric (1) from a Galilean metric depends only on $f \equiv \gamma_{00}$, which should be close to 1 for a weak field, and since Eq. (1) is valid only for $r > R$, an obvious (and well-known) definition is:

$$\lambda' \equiv r_g/R \ll 1 \qquad (r_g \equiv 2\, GM/c^2).$$

In the same way as the special-relativistic energy of a point particle can be separated, if its velocity is small as compared with $c$, into rest-mass energy, Newtonian kinetic energy, and potential energy, we should be able to make a such kind of separation for the field $\mu^*$. In a further step, we could then insert (4) or (9) into (3) and thus obtain an approximation for $M$, which also would be separated into well-identified Newtonian terms. This separation occurs naturally if we have a framework suitable for asymptotic approximations, namely if we are able to consider a *family* of massive bodies ($B_\lambda$), depending on a small parameter $\lambda$ – as is needed to give a mathematical meaning to the assumption $\lambda \ll 1$. A such framework has been set for a general system of extended bodies in the case of the scalar theory [11]. In GR, two such frameworks were previously proposed, one by Futamase & Schutz [7] and the other by Rendall [8], apparently without later developments in celestial mechanics of extended bodies. The approach of Futamase & Schutz [7] contains essential elements of the two later approaches, though it assumes a restrictive initial condition for the metric field [their equations (3.13)$_{4-5}$]. Rendall's mathematical analysis [8] is general, but it starts from an *a priori* given family of solutions to the gravitational equations, without attempting to relate the construction of this family to the physically given gravitational system.

In the approach followed by the author [11], as in the two previous ones, one considers a family of matter and gravitational fields [thus a family ($S_\lambda$) of gravitating systems], indexed by the parameter $\lambda$, that becomes small for a weak field: this is appropriate to derive asymptotic expansions as $\lambda \to 0$. [2] As in Futamase & Schutz [7], the family of fields is defined by a family of initial data for them. We start from the physically given, weakly gravitating system S, of a general nature (*e.g.* the solar system), which we are really interested in, and which should correspond to a given, small value $\lambda_0$ for the parameter $\lambda$, and we deduce a family of initial data from the initial data that applies to the given system S [11]. To do this, we use the *NL* (Newtonian limit) *condition* according to which, in the limit of a weakly gravitating system ($\lambda \to 0$), the matter fields, as well as the gravitational field produced by it, both become closer and closer to Newtonian. This condition can be made precise, because there is an *exact similarity transformation in Newtonian gravity* [7, 11]. If we have the following fields: pressure $p^1$, density $\rho^1 = F_1(p^1)$, Newtonian potential $U_N^1$ and velocity $\mathbf{u}^1$, obeying the continuity equation, Poisson's equation, and Euler's equation, then the fields $p^\lambda(\mathbf{x}, t) = \lambda^2\, p^1(\mathbf{x}, \sqrt{\lambda}\, t)$, $\rho^\lambda(\mathbf{x}, t) = \lambda \rho^1(\mathbf{x}, \sqrt{\lambda}\, t)$, $U_N^\lambda(\mathbf{x}, t) = \lambda U_N^1(\mathbf{x}, \sqrt{\lambda}\, t)$, and $\mathbf{u}^\lambda(\mathbf{x}, t) = \sqrt{\lambda}\, \mathbf{u}^1(\mathbf{x}, \sqrt{\lambda}\, t)$, also obey these equations [provided the state equation for system $S_\lambda$ is $F_\lambda(p^\lambda) = \lambda F_1(\lambda^{-2} p^\lambda)$], for any $\lambda > 0$. Due to this similarity transformation, the NL condition imposes the orders in $\lambda$ of the fields as $\lambda \to$ 0, at least for the matter fields, because they are common to Newtonian and relativistic gravity. To ensure that the orders of the fields are those given above, it is natural to define the initial data of system $S_\lambda$ for matter fields by *applying the Newtonian similarity transformation to the initial data,* either that of system $S_{\lambda=1}$ [7], or that of the physically given system $S = S_{\lambda_0}$ [11]. Moreover, the orders of the fields are such that, *when one changes the time and mass units for system $S_\lambda$, multiplying the starting time unit by $\lambda^{-1/2}$ and the mass unit by $\lambda$, then $\lambda$ becomes proportional to $1/c^2$ and all matter fields are of order zero with respect to $\lambda$* [11]. This is immediate to check and applies to the scalar theory and to GR as well. Therefore, the derivation of asymptotic expansions becomes elementary. After the asymptotic expansions of the different fields have been written in the varying units and equations have been derived for them, the same expansions and equations remain valid in the starting units which are independent of the small parameter $\lambda$ – up to the fact that the order of the coefficient fields is changed [9]. The method of asymptotic expansions provides thus a straightforward approximation method for weak gravitational fields, which is close to the standard method to obtain PN expansions for extended bodies, first introduced by Fock [15] – but with two differences: i) the matter fields also are expanded, and ii) since the expansion is based on a conceptual family of systems, defined by a family of initial conditions, one must use these initial conditions to obtain fully explicit expansions [11].

---

[2] In this context, all fields (including the matter fields) have to be expanded with respect to $\lambda$, not merely the gravitational field as is done in the standard PNA. Indeed, if one keeps some fields "unexpanded", then one should not separate equations corresponding to different orders in $\lambda$, with the result that there are not enough equations. This is illustrated for the scalar theory in Ref. [11].

Let us do this to get the coefficients in the expansion of the field source in GR, Eqs. (4) and (6). We write first-order expansions in the small parameter $1/c^2$ (using the varying units):

$$\rho^* = \rho_0 + \rho^*_1/c^2 + O(1/c^4), \tag{12}$$

$$\mu^* = \mu^*_0 + \mu^*_1/c^2 + O(1/c^4), \tag{13}$$

*etc*. At the initial time, *i.e.* that of the initial conditions, this must be equal to the initial data:

$$\rho_0(\mathbf{x}, t=0) + \rho^*_1(\mathbf{x}, t=0)/c^2 + O(1/c^4) = \rho_{\text{data}}(\mathbf{x}, t=0). \tag{14}$$

This implies the following:

$$\rho_0(\mathbf{x}, t=0) = \rho_{\text{data}}(\mathbf{x}, t=0), \qquad \rho^*_1(\mathbf{x}, t=0) = 0. \tag{15}$$

On the other hand, we get from (6):

$$\mu^*_0 = \rho_0, \quad \mu^*_1 = \rho^*_1 + \rho_0 \Pi_0, \tag{16}$$

and using now the condition that the situation is static, we obtain by (15):

$$\mu^*_1 = \rho_0 \Pi_0. \tag{17}$$

Hence, we have the expansion of the active mass of Schwarzschild's solution in GR, Eqs. (3)-(4):

$$M = M_0 + M_1/c^2 + O(1/c^4), \qquad M_0 = \int \rho_0 \, dV, \qquad M_1 = \int \rho_0 \Pi_0 \, dV. \quad \text{(GR)} \tag{18}$$

But the zero-order fields obey the equations of Newtonian gravity [7-8]. Thus $M_0$ is the Newtonian mass and $M_1$ is the Newtonian elastic energy, whose expression is easily found from Eqs. (74.03), (74.07), (74.24) and (76.06) of Fock [4]:

$$M_1 = (5/3)\varepsilon_B \qquad \varepsilon_B \equiv \int \rho_0 \, u_B \, dV/2, \qquad u_B(r) \equiv \int_r^\infty GM_0(r')\, dr'/r'^2. \tag{19}$$

(Recall that here the velocity field is zero.) Note that the expansions $(16)_1$, (17) and (18) are just those that one obtains by entering in Eq. (6) the obvious approximations

$$\rho^* \approx \rho_0, \qquad \Pi \approx \Pi_0, \qquad \rho^*\Pi/c^2 \ll \rho^*. \qquad (20)$$

Thus, the expansion obtained for the active mass has nothing surprising. In the scalar theory, a similar way of reasoning leads, using Eq. $(2.24)_2$ of Ref. [9], to

$$M = M_0 + M_1/c^2 + O(1/c^4), \quad M_0 = \int \rho_0 \, dV, \quad M_1 = M^1 + \int \rho_0(\Pi_0 + u_B) \, dV \quad \text{(scalar theory)}, \qquad (21)$$

where $M^1$ is defined by Eqs. (3.6)-(3-7) of Ref. [9], and this leads in the same way to

$$M_1 = (17/3)\varepsilon_B \quad \text{(scalar theory)}, \qquad (22)$$

instead of $(19)_1$.

Thus, the active mass in Schwarzschild's metric admits a first-order PN expansion, in which the first term is the Newtonian mass, and the second term is the elastic energy – augmented, in the scalar theory, by a term arising from the fact that the gravitational field reinforces its own source.

*2.3 PN equations of motion for a massive test particle: standard and asymptotic version*

The exact geodesic equations of motion for a test particle can be rewritten in terms of the coordinate time $t$ and the corresponding velocity $u^i = dx^i/dt$ [1, Eq. (9.1.2)]. In the case of any static gravitational field, it is easy to check that they in turn can be rewritten as follows [14, Eqs. (19) and (20) with $f_{,0} = 0$ and $g_{ij,0} = 0$]:

$$\frac{du^i}{dt} = \frac{f_{,j}}{f} u^j u^i - \Gamma^i_{jk} u^j u^k - \frac{c^2}{2} g^{ij} f_{,j}, \qquad f \equiv \gamma_{00}, \qquad (23)$$

with $\Gamma^i_{jk} = g^{im} \Gamma_{mjk}$ the second-kind Christoffel symbols of the spatial metric $\mathbf{g} = (g_{ij})$ in the considered coordinates, adapted to the static reference frame,[3] and where $(g^{ij})$ is the inverse of matrix $(g_{ij})$. The last term in Eq. (23) is a space vector, which is easy to evaluate for Schwarzschild's metric (1). *Adopting henceforth the "Cartesian" coordinates (2),* we get thus for the static case with spherical symmetry:

$$\frac{du^i}{dt} = \frac{f_{,j}}{f} u^j u^i - \Gamma^i_{jk} u^j u^k - \frac{c^2}{2} f f_{,i}, \qquad (24)$$

where, the particle having to remain outside the body, we have[4] by Eq. (1):

$$f(\mathbf{x}) = 1 - 2\,GM/(c^2 r), \qquad r \equiv (\mathbf{x}.\mathbf{x})^{1/2} > R. \qquad (25)$$

---

[3] An adapted time coordinate $t$ is also selected, and this actually imposes the time coordinate up to a change of time unit. Thus $f \equiv \gamma_{00}$ is uniquely defined, also in GR or in any other generally-covariant theory.

[4] Henceforth, we use the scalar product defined by $\mathbf{x}.\mathbf{y} \equiv x^i.y^i$ in the coordinates (2).

To obtain PN expansions of this equations, we again just need to take $1/c^2$ as the small parameter and to consider that all matter fields (hence also the active mass $M$) are of order zero, since this is equivalent to writing expansions in terms of the true small parameter $\lambda$, after having multiplied the starting time unit by $\lambda^{-1/2}$ and the mass unit by $\lambda$. Naturally, the velocity $\mathbf{u}$ of the test particle also depends on the small parameter. In order to keep consistency with the Newtonian limit, we must admit that (in fixed units), the velocity is of the order of $\lambda^{1/2}$ and hence it also is of order zero in the new units. The expansion of the Christoffel symbols follows from writing successively

$$g_{ij} = \delta_{ij} + \left(\frac{1}{f} - 1\right) n_i n_j, \qquad \mathbf{n} \equiv \mathbf{x}/\|\mathbf{x}\|, \tag{26}$$

$$g_{ij} = \delta_{ij} + (2\,GM/(c^2 r))\, n_i n_j + O(1/c^4), \tag{27}$$

$$\Gamma_{ijk} \equiv \frac{1}{2}(g_{ij,k} + g_{ik,j} - g_{jk,i}) = 2\frac{GM}{c^2}\Gamma'_{ijk} + O(1/c^4), \quad g'_{ij} \equiv \frac{n_i n_j}{r}.$$

The distinction between $M_0$ and $M$ will affect the two first terms on the r.h.s. of Eq. (24) only by $O(1/c^4)$ corrections, which are neglected at the first PN (1PN) approximation. Indeed we get

$$\frac{f_{,j}}{f} = f_{,j} + O\left(\frac{1}{c^4}\right) = 2\frac{GM}{c^2}\frac{n_j}{r^2} + O\left(\frac{1}{c^4}\right), \tag{28}$$

$$\Gamma^i_{jk} u^j u^k = 2\frac{GM}{c^2}\left(\mathbf{u}^2 - \frac{3}{2}(\mathbf{u}.\mathbf{n})^2\right)\frac{n_i}{r^2} + O\left(\frac{1}{c^4}\right). \tag{29}$$

The last term in Eq. (24) is $O(1)$, it has the exact expression:

$$-\frac{c^2}{2} f\, f_{,i} = -\left(1 - \frac{2GM}{c^2 r}\right)\frac{GM}{r^2} n_i, \tag{30}$$

and for this term the distinction between $M_0$ and $M$ will make an $O(1/c^2)$ difference.

At this stage, we have two possible choices to obtain 1PN equations of motion, correct up to $O(1/c^2)$ terms included. First, we may insert (28)-(30) into the exact equation (24). We get thus

$$\frac{d\mathbf{u}}{dt} = \frac{GM}{r^2}\left\{\left[-1 + \frac{2}{c^2}\left(\frac{GM}{r} + \frac{3}{2}(\mathbf{u}.\mathbf{n})^2 - \mathbf{u}^2\right)\right]\mathbf{n} + \frac{2}{c^2}(\mathbf{u}.\mathbf{n})\mathbf{u}\right\} + O\left(\frac{1}{c^4}\right) \tag{31}$$

and, if we want to directly obtain a useful equation, we are led to neglect the $O(1/c^4)$ term, thus taking $M_{(1)} \equiv M_0 + M_1/c^2$ for the active mass and defining approximate solutions $\mathbf{x}_s$ and $\mathbf{u}_s$ (where index s means "standard") for the position and velocity of the test particle:

$$\frac{d\mathbf{x}_s}{dt} = \mathbf{u}_s, \quad \frac{d\mathbf{u}_s}{dt} = \frac{GM_{(1)}}{r^2}\left\{\left[-1 + \frac{2}{c^2}\left(\frac{GM_{(1)}}{r} + \frac{3}{2}(\mathbf{u}_s.\mathbf{n})^2 - \mathbf{u}_s^2\right)\right]\mathbf{n} + \frac{2}{c^2}(\mathbf{u}_s.\mathbf{n})\mathbf{u}_s\right\}, \tag{32}$$

with the same initial conditions as for the exact solution, *i.e.*

$$\mathbf{x}_s(t_0) = \mathbf{x}(t_0) \equiv \mathbf{x}_{00}, \qquad \mathbf{u}_s(t_0) = \mathbf{u}(t_0) \equiv \mathbf{u}_{00}. \tag{33}$$

Equation (32) is the equation (3.1.46) of Brumberg [16], for the relevant case (GR in "standard coordinates", according to an usual terminology). Thus Eq. (32) is just what corresponds to the standard PN approximation (developed for extended bodies), for the particular case of a test particle in a weak Schwarzschild field – and we shall name it the *standard equation*. The second possibility is to search for an *asymptotic expansion* of the position **x**(*t*) and the velocity **u**(*t*) of the test particle, thus writing

$$\mathbf{x} = \mathbf{x}_0 + \mathbf{x}_1/c^2 + O(1/c^4), \qquad \mathbf{u} = \mathbf{u}_0 + \mathbf{u}_1/c^2 + O(1/c^4), \qquad (34)$$

and to insert this in Eq. (31). Since each of the two vector unknowns **x** and **u** is now splitted into two parts: a zero-order part and a first-order (1PN) correction, we must also split Eq. (31) into two separate vector equations, and thus we must use asymptotic expansions (whose coefficients are, by definition, *independent* of the small parameter) for each quantity entering Eq. (31) – in particular we must use the expansion (18) [or (21)] for the active mass *M*. This gives

$$\frac{d\mathbf{u}_0}{dt} = -\frac{GM_0}{r_0^2}\mathbf{n}_0, \qquad (\mathbf{n}_0 \equiv \mathbf{x}_0/\|\mathbf{x}_0\|, \quad r_0 \equiv \|\mathbf{x}_0\|) \qquad (35)$$

$$\frac{d\mathbf{u}_1}{dt} = \frac{GM_0}{r_0^2} \times \left\{ \left[ \left( \frac{2GM_0}{r_0} - \frac{M_1}{M_0} + 3\left[ (\mathbf{u}_0 \cdot \mathbf{n}_0)^2 + \frac{\mathbf{x}_0 \cdot \mathbf{x}_1}{r_0^2} \right] - 2\mathbf{u}_0^2 \right) \right] \mathbf{n}_0 + 2(\mathbf{u}_0 \cdot \mathbf{n}_0)\mathbf{u}_0 - \frac{\mathbf{x}_1}{r_0} \right\}. \qquad (36)$$

Note that, in contrast to Eq. (32), the equations (35) and (36) of the asymptotic method have the status of exact equations [although they define merely a part of the relevant exact solution, Eq. (34)]. Of course, Eqs. (35)-(36) must be completed by the following consequence of Eqs. (34)$_{1-2}$ (under the assumption that the expansion (34)$_1$ has sufficient uniformity with respect to the time *t*):

$$\frac{d\mathbf{x}_0}{dt} = \mathbf{u}_0, \qquad \frac{d\mathbf{x}_1}{dt} = \mathbf{u}_1, \qquad (37)$$

and by the initial conditions, which are obtained by inserting the asymptotic expansions (34) into the initial conditions (33) for the exact solution:

$$\mathbf{x}_0(t_0) = \mathbf{x}_{00}, \qquad \mathbf{x}_1(t_0) = \mathbf{0}, \qquad (38)$$

$$\mathbf{u}_0(t_0) = \mathbf{u}_{00}, \qquad \mathbf{u}_1(t_0) = \mathbf{0}. \qquad (39)$$

*2.4 Numerical test for Mercury*

As we have just seen, in the case of a massive test particle moving along a geodesic line of Schwarzschild's metric, there are two different systems of equations for the first PN approximation: the "standard" one, Eqs. (32)$_{1-2}$, and the "asymptotic" one, Eqs. (35)-(37). Both systems are correct if one neglects $O(1/c^4)$ terms, hence both should give nearly identical results in the solar system, up to possible numerical problems. In order to check this, the solutions of both systems were numerically investigated and compared, in the case of Mercury. To this aim, the two differential systems: the *"standard"* one, *i.e.* Eq. (32) with initial conditions (33), and the *"asymptotic"* one, *i.e.* Eqs. (35)-(37) with initial conditions (38)-(39), were solved by using the Matlab routine ODE113, a variable order solver with internally adjusted, variable step size. (The high numerical performance of this routine for solving celestial-mechanical problems was previously checked on the Newtonian *N*- bodies problem, see Ref. [17]. The same numerical tolerances were selected in the ODE113 routine here as those found optimal in the latter work, *i.e.* RelTol = 5×10$^{-13}$ and AbsTol = 10$^{-15}$.) The initial conditions were selected as follows. The osculating Keplerian elements at the initial time were taken to be: semi-major axis $a$ = 0.387098356374731 au, eccentricity $e$ = 0.205630584112448, inclination $i$ = 0, mean longitude $\lambda$ = 0, longitude of node $\Omega$ = 0, longitude of perihelion $\varpi \equiv \Omega + \omega = 0$. The two first values were calculated from the DE403 ephemeris for the Julian Day 2451600.5, *i.e.* 26 February 2000 at 0H00 "Temps Terrestre", and the four latter values mean simply that the orbital plane was taken as the reference plane, with the *x* axis coinciding with the direction of the perihelion, and with the planet being at perihelion at the initial time − this was not actually true for Mercury at the above date, but this is immaterial here: all perturbations by the planets being neglected, the calculation is merely a numerical test, hence the time origin may be changed. Then the initial conditions for **x** and **ẋ** were obtained by coming back from osculating elements to rectangular coordinates (the author's routines for these conversions have been checked in Ref. [17]), and were found to be (in au and au/d, respectively):

$$\mathbf{x}_{00} = (0.307499095244427, 0, 0), \qquad \mathbf{u}_{00} = (0, 0.0340617258068722, 0). \qquad (40)$$

The constant of gravitation was taken, in the International Astronomical Units used, at the IERS 1992 value, *i.e.* $G$ = (0.01720209895)$^2$ = 2.95912208285591×10$^{-4}$ au$^3$/d$^2$, and the mass of the Sun is the unity in the IAU system. For this calculation, it was considered that the mass which is equal to one is the zero-order mass, thus $M_0$ = 1 (see the comments at the end of this Section). Finally, the calculation was done in the *GR case,* thus $M_1$ = (5/3)$\varepsilon_B$ [Eq. (19)$_1$], and it was found that $\varepsilon_B \approx 0.106$ (IAU) by doing numerically the two quadratures implied in Eq. (19)$_3$, taking the density profile of the Sun as given by Bakouline *et al.* [18]. This gives $M_1/c^2 \approx 5.88 \times 10^{-6}$ solar mass.

The time evolution of the distance between the heliocentric positions of Mercury, as calculated either with the standard method or the asymptotic method:

$$\delta R(t) \equiv \| \mathbf{x}_s(t) - (\mathbf{x}_0 + \mathbf{x}_1/c^2)(t) \| \quad (\text{au}), \qquad (41)$$

was calculated for one century. It was found that the distance increases quite quickly: a log-log diagram shows that it increases just like $(t - t_0)^2$ (Fig. 1). Now, *in the present case* of a test particle in a weak Schwarzschild field, the standard approximation, Eq. (32), is certainly closer to the exact solution, because it is derived from Eq. (31) just by neglecting the $O(1/c^4)$ term, whereas the asymptotic approximation introduces the expansions (34), which in the present context merely add a further approximation. To obtain accurate predictions using the asymptotic method, it is necessary to "reinitialize". By this, we mean that, after some time $\delta t$ is elapsed since the initial time $t_0$, one writes a new asymptotic expansion (34), with initial conditions like (38)-(39), but with $t'_0 = t_0 + \delta t$ as the initial time. Yet the exact solution is not known at time $t'_0$, hence the best one can do is to take the initial conditions for the new expansion (denoted with primes) as the values of the former expansion at time $t'_0$:

$$\mathbf{x'}_0(t'_0) = \mathbf{x'}_{00} \equiv (\mathbf{x}_0 + \mathbf{x}_1/c^2)(t'_0), \qquad \mathbf{x'}_1(t'_0) = \mathbf{0}, \qquad (42)$$

and the like for the velocity. At time $t''_0 = t'_0 + \delta t$, one starts a third expansion, *etc.*

Figures 2 and 3 show the time evolution, respectively over 2000 days and one century, of the distance between the "standard" position [that is, the numerical solution of Eqs. (32) with initial conditions (33)] and the "reinitialized asymptotic" position − that is, the position obtained by numerically solving, successively, Eqs. (35)-(36) with initial conditions (38)-(39) in the interval $[t_0, t_0 + \delta t]$, then Eqs. (35)-(36) with initial conditions (42) in the interval $[t'_0, t'_0 + \delta t]$, *etc.* On each of Figs. 1 and 2, four different values of the reinitialization interval $\delta t$ are compared: $\delta t = \infty$ (no reinitialization), $\delta t = 40$ days, $\delta t = 10$ days, and $\delta t = 4$ days. It can be seen that the error depends very sensitively on $\delta t$. For $\delta t = 40$ days (and even more for larger values), the reinitialization has a negative effect. With $\delta t = 4$ days, the difference between the asymptotic method and the standard one (hundred kilometers, or some 0.4", after one century) is small enough, so that the asymptotic method can be used for testing a theory of gravitation. The reinitialization was expected to increase marginally the computation time, but actually it increases significantly (*ca.* five times with $\delta t = 4$ days), for a contingent reason: the efficient ODE solvers are of a high order and, seemingly for this reason, need a lot of very small steps before reaching the large step size of the "established regime". Here the latter large step size cannot be reached, because a new call of the solver is done before this.

A last comment is that, for a complete calculation including the Newtonian perturbations of the planets, using the IAU system and those masses of the planets which are recommended by astronomical organizations, that mass of the Sun which is to be taken equal to one turns out to be the total active mass of the 1PN approximation, thus $M_{(1)} \equiv M_0 + M_1/c^2 = 1$ instead of $M_0 = 1$. The reason is that the current ephemerides are based on the standard PN approximation and do not consider zero-order masses. This has been checked by numerical calculations. But the mass correction $M_1/c^2 \approx 5.88 \times 10^{-6}$ solar mass is obviously too small to affect the magnitude of the difference, at given values of $M_0$ and $M_1$, between the standard and asymptotic PN approximations.

## 3. Conclusion

In this paper, the recently developed "asymptotic method" of PN expansion has been compared with the standard method, in the case of a massive test particle moving along a geodesic line of a weak Schwarzschild field. In this particular, but practically important case, both methods are equally justified from the theoretical viewpoint. The asymptotic method has a practical weakness, in that the test particle is followed on its trajectory by two different positions: the zero-order position $\mathbf{x}_0$ and the 1PN position, $\mathbf{x}_{(1)} \equiv \mathbf{x}_0 + \mathbf{x}_1/c^2$ (this is mathematically necessary to get asymptotic expansions). At the initial time, these two positions coincide but, as the time flows, they go further and further from one another. Thus, the Newtonian attraction (34) refers to a position which differs more and more from the 1PN position, the latter being supposed to be more correct. This is clearly a detrimental feature, for the starting equation for PN approximation, Eq. (31), assigns one and only position to the particle for the zero-order part of the acceleration and its first-order correction.

On the other hand, in the more general case of a system of several extended bodies, unlike in the present case, we cannot start from the exact solution (25) for the gravitational field, and also we do not have exact equations of motion for the mass centers. Hence the most natural thing to do in the general case seems to be setting asymptotic expansions for all fields so as to obtain separate equations, and then integrating the local equations in order to obtain equations for the mass centers – that is, using the asymptotic method. The weak point of the asymptotic method, which is the error increase in $(t - t_0)^2$, may be easily accommodated by "reinitializing", *i.e.,* by substituting $t_0 + \delta t$, then $t_0 + 2\delta t$, *etc.,* with $|\delta t|$ sufficiently small, for the initial time $t_0$ in the differential system that governs the motion of the mass centers.

**Acknowledgement.** I am grateful to P. Bretagnon, C. Marchal and P. Teyssandier for helpful remarks. The initial data taken from the DE403 ephemeris was read from the web site of the Institut de Mécanique Céleste du Bureau des Longitudes, Paris.

# References

[1] Weinberg S., *Gravitation and Cosmology* (J. Wiley & Sons, New York) 1972.
[2] Newhall X. X., Standish E. M. and Williams J. G., *Astron. & Astrophys.,* **125**, 150-167 (1983).
[3] Lestrade J.-F. and Bretagnon P., *Astron. & Astrophys.,* **105**, 42-52 (1982).
[4] Fock V., *The Theory of Space, Time and Gravitation* (2nd English Edition, Pergamon, Oxford) 1964.
[5] Chandrasekhar S., *Astrophys. J.,* **142** (1965) 1488-1512.
[6] Kevorkian J. and Cole J. D., *Perturbation Methods in Applied Mathematics* (Springer, New York - Heidelberg - Berlin) 1981.
[7] Futamase T. and Schutz B. F., *Phys. Rev. D,* **28**, 2363-2372 (1983).
[8] Rendall A. D., *Proc. R. Soc. Lond. A,* **438**, 341-360 (1992).
[9] Arminjon M., *Roman. J. Phys.,* **45**, N° 9-10 (2000), Paper I. See also preprint astro-ph/0006093.
[10] Arminjon M., *Roman. J. Phys.,* **45**, N° 9-10 (2000), Paper II. See also preprint astro-ph/0006093.
[11] Arminjon M., *Roman. J. Phys.,* **45**, N° 5-6 (2000). Also as preprint gr-qc/0003066.
[12] Thirring W., *Classical Field Theory* (2nd English Edition, Springer, New York - Wien) 1986, p. 215.
[13] Wald R. M., *General Relativity* (The University of Chicago Press, Chicago - London) 1984, p. 126.
[14] Arminjon M., *Rev. Roum. Sci. Tech.- Méc. Appl.,* **43**, 135-153 (1998). Also as preprint gr-qc/9912041.
[15] Fock V., *J. Phys. USSR,* **1**, 89-116 (1939) [*Zh. Eksp. Teor. Fiz.,* **9**, 375- (1939)].
[16] Brumberg V. A., *Essential Relativistic Celestial Mechanics* (Iop Publishing/Adam Hilger, Bristol) 1991.
[17] Arminjon M., preprint astro-ph/0105217 (2001).
[18] Bakouline P., Kononovitch E. and Moroz V.I., *Astronomie Générale* (Mir, Moscow) 1975, Pp. 102-105 (Russian 3rd Edition: Kurs Obshcheï Astronomiï, Izd. Nauka, 1973).

# Figure Captions

**Figure 1.** Log-Log diagram for the time evolution of the distances between the heliocentric positions of Mercury, calculated either with the standard PN equations (32)-(33) or the asymptotic PN equations (35)-(39). The slope of the median line is close to 2.

**Figure 2.** Distance between the heliocentric positions of Mercury, calculated with either the standard PN equations (32)-(33) or the asymptotic PN equations (35)-(39), over a time span of 2000 days; the four plots are for different values of the reinitialization time in the asymptotic method: $\delta t = \infty$ (no reinitialization), $\delta t = 40$ days, $\delta t = 10$ days, and $\delta t = 4$ days.

**Figure 3.** As for Fig. 2, but here over a time span of 36520 days (one century = 36525 days).

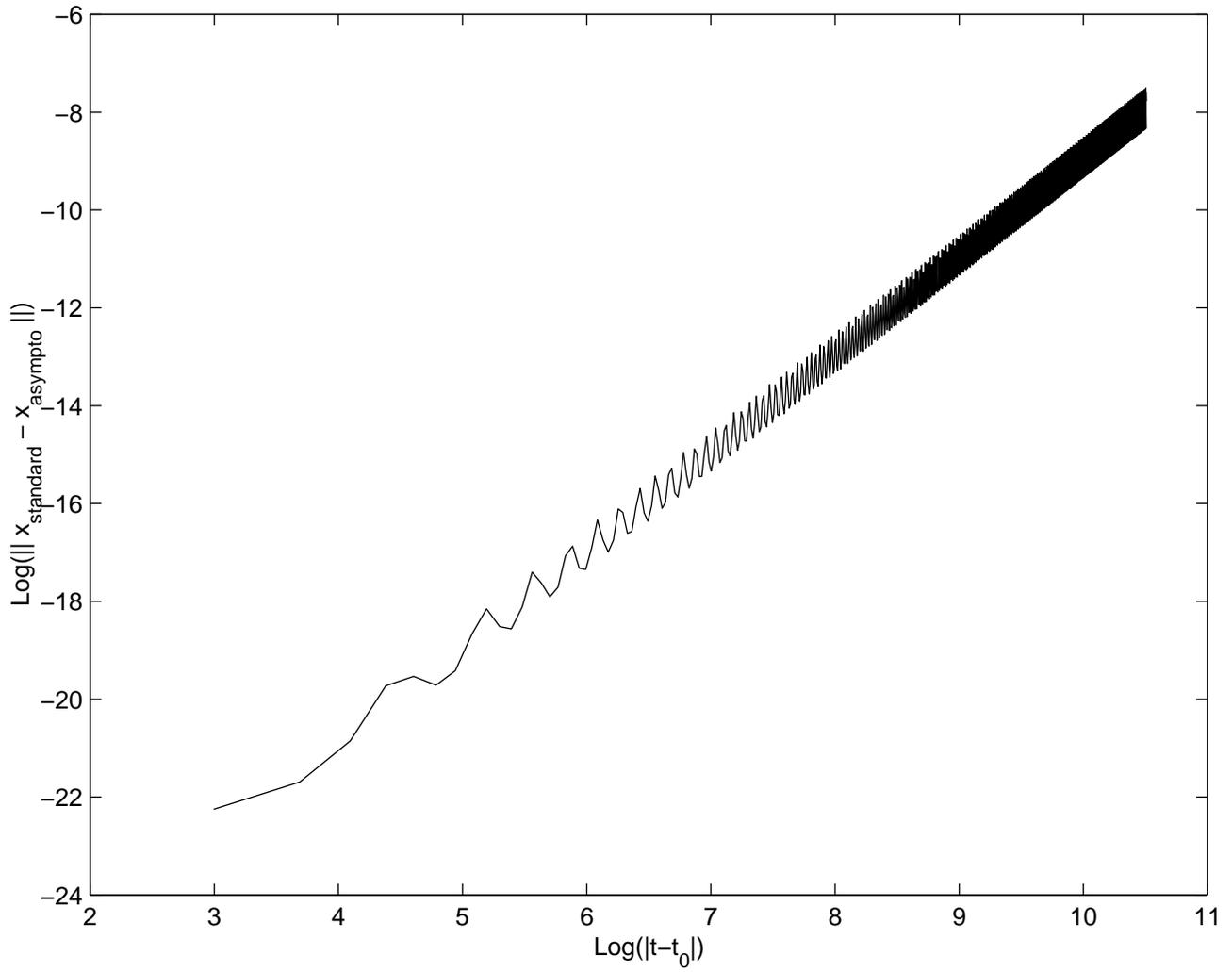

Difference between standard and asymptotic PN calculations for Mercury

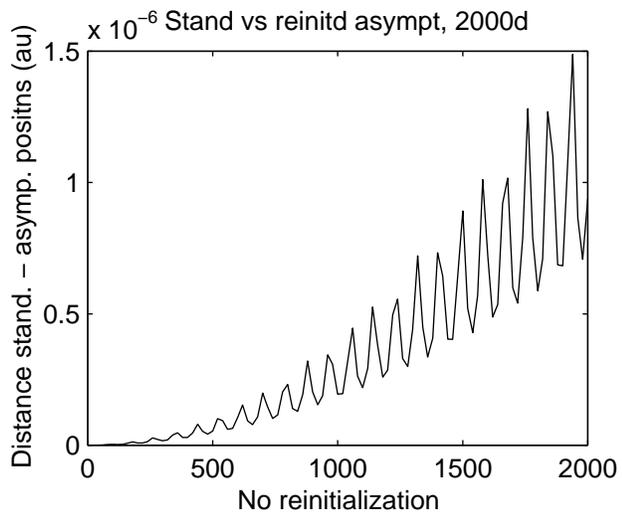
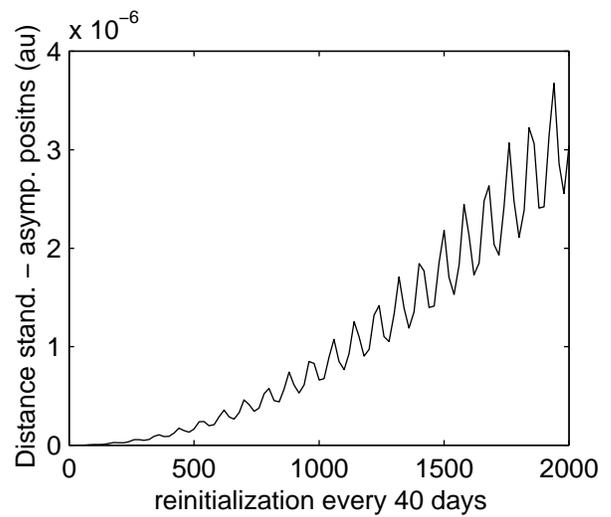
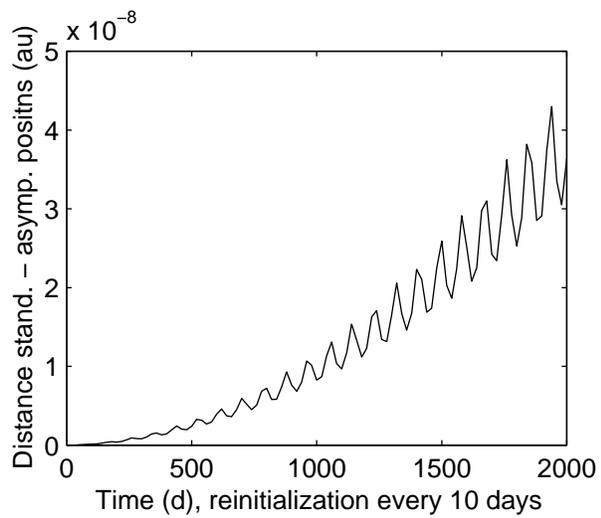
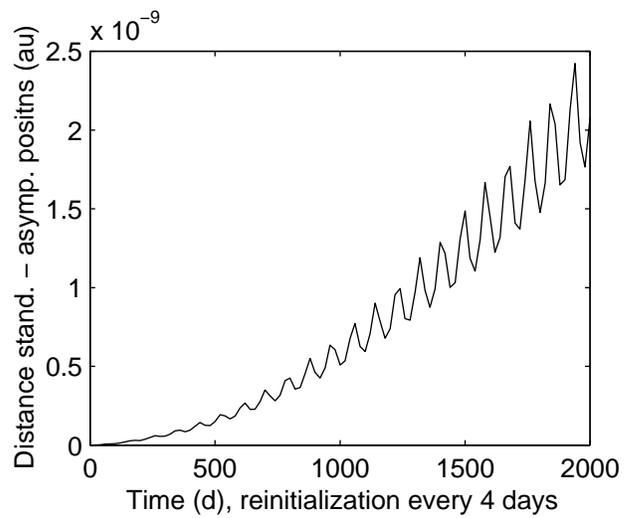

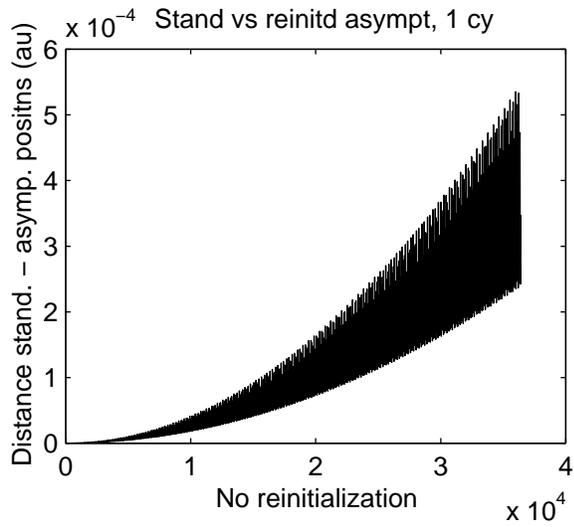
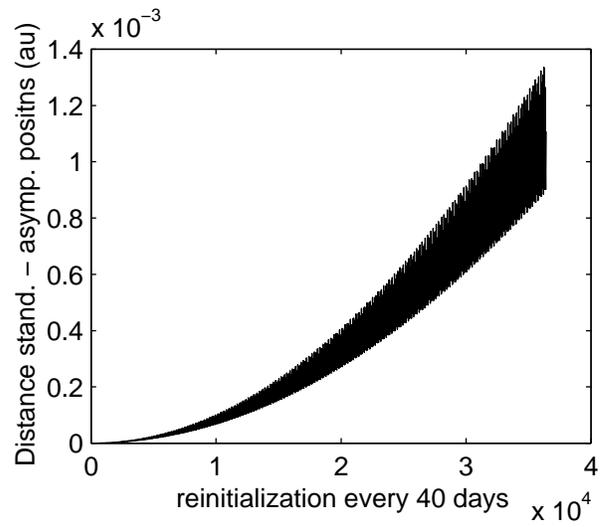
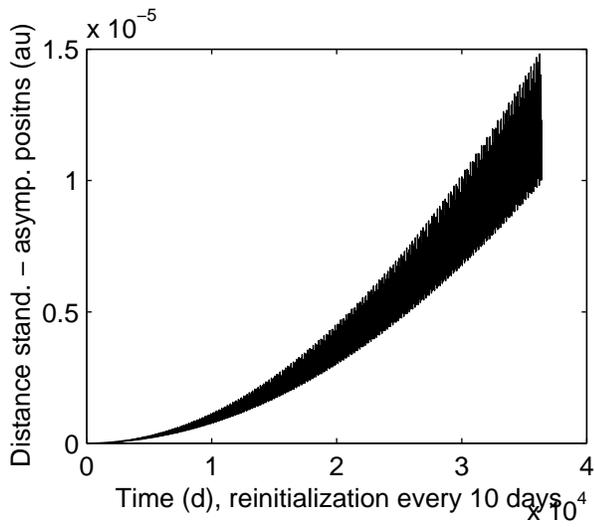
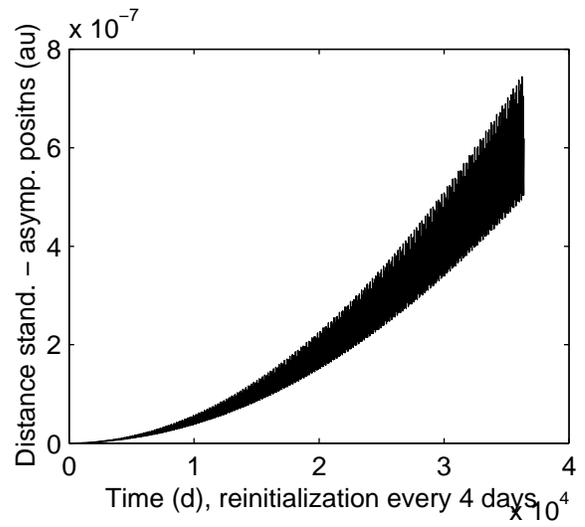